\documentstyle[12pt,epsf]{article}
\expandafter\ifx\csname amssym12.def\endcsname\relax \else\endinput\fi
\expandafter\edef\csname amssym12.def\endcsname{%
       \catcode`\noexpand\@=\the\catcode`\@\space}
\catcode`\@=11
\def\undefine#1{\let#1\undefined}
\def\newsymbol#1#2#3#4#5{\let\next@\relax
 \ifnum#2=\@ne\let\next@\msafam@\else
 \ifnum#2=\tw@\let\next@\msbfam@\fi\fi
 \mathchardef#1="#3\next@#4#5}
\def\mathhexbox@#1#2#3{\relax
 \ifmmode\mathpalette{}{\m@th\mathchar"#1#2#3}%
 \else\leavevmode\hbox{$\m@th\mathchar"#1#2#3$}\fi}
\def\hexnumber@#1{\ifcase#1 0\or 1\or 2\or 3\or 4\or 5\or 6\or 7\or 8\or
 9\or A\or B\or C\or D\or E\or F\fi}
\font\tenmsa=msam10 scaled\magstep1
\font\sevenmsa=msam7 scaled\magstep1
\font\fivemsa=msam5 scaled\magstep1
\newfam\msafam
\textfont\msafam=\tenmsa
\scriptfont\msafam=\sevenmsa
\scriptscriptfont\msafam=\fivemsa
\edef\msafam@{\hexnumber@\msafam}
\mathchardef\dabar@"0\msafam@39
\def\dashrightarrow{\mathrel{\dabar@\dabar@\mathchar"0\msafam@4B}}
\def\dashleftarrow{\mathrel{\mathchar"0\msafam@4C\dabar@\dabar@}}

\def\ulcorner{\delimiter"4\msafam@70\msafam@70 }
\def\urcorner{\delimiter"5\msafam@71\msafam@71 }
\def\llcorner{\delimiter"4\msafam@78\msafam@78 }
\def\lrcorner{\delimiter"5\msafam@79\msafam@79 }
\def\yen{{\mathhexbox@\msafam@55 }}
\def\checkmark{{\mathhexbox@\msafam@58 }}
\def\circledR{{\mathhexbox@\msafam@72 }}
\def\maltese{{\mathhexbox@\msafam@7A }}
\font\tenmsb=msbm10 scaled\magstep1
\font\sevenmsb=msbm7 scaled\magstep1
\font\fivemsb=msbm5 scaled\magstep1
\newfam\msbfam
\textfont\msbfam=\tenmsb
\scriptfont\msbfam=\sevenmsb
\scriptscriptfont\msbfam=\fivemsb
\edef\msbfam@{\hexnumber@\msbfam}

\def\widehat#1{\setbox\z@\hbox{$\m@th#1$}%
 \ifdim\wd\z@>\tw@ em\mathaccent"0\msbfam@5B{#1}%
 \else\mathaccent"0362{#1}\fi}
\def\widetilde#1{\setbox\z@\hbox{$\m@th#1$}%
 \ifdim\wd\z@>\tw@ em\mathaccent"0\msbfam@5D{#1}%
 \else\mathaccent"0365{#1}\fi}
\font\teneufm=eufm10 scaled\magstep1
\font\seveneufm=eufm7 scaled\magstep1
\font\fiveeufm=eufm5 scaled\magstep1
\newfam\eufmfam
\textfont\eufmfam=\teneufm
\scriptfont\eufmfam=\seveneufm
\scriptscriptfont\eufmfam=\fiveeufm

\csname amssym.def\endcsname
\newif{\ifcomentarios}
\comentariosfalse

\newcommand{\be}{\begin{equation}}
\newcommand{\ee}{\end{equation}}
\newcommand{\bma}{\begin{displaymath}}
\newcommand{\ema}{\end{displaymath}}
\newcommand{\bc}{\begin{center}}
\newcommand{\ec}{\end{center}}
\newcommand{\text}{\rm}

\newcommand{\uflex}
{{\scriptstyle {\raise 9pt\hbox{$\backslash$}\,\!\!\!\!\!\Bigg\vert}}}
\newcommand{\ncm}{\newcommand}

\ncm{\rncm}{\renewcommand}
\ncm{\id}{{\bf 1}}
\ncm{\beq}{\begin{equation}}
\ncm{\eeq}{\end{equation}}
\ncm{\ba}{\begin{array}}
\ncm{\bea}{\begin{eqnarray}}
\ncm{\beanon}{\begin{eqnarray*}}
\ncm{\ea}{\end{array}}
\ncm{\eea}{\end{eqnarray}}
\ncm{\eeanon}{\end{eqnarray*}}
\ncm{\fns}{\footnotesize}
\newcounter{eqnr}
\newenvironment{eqnarrayabc}{\stepcounter{equation}
\setcounter{eqnr}{\value{equation}}\setc{0}
  \rncm{\theequation}{\thesection.\arabic{eqnr}\alph{equation}}
  \begin{eqnarray}}{\end{eqnarray}\setc{\value{eqnr}}}
\ncm{\eqboxabc}[3]{\newline\parbox[t]{1.5cm}{#1}\hfill
  \parbox[b]{12cm}{\begin{eqnarray*} #3\end{eqnarray*}}\hfill
   \parbox[b]{1.5cm}{\vspace{-0.0cm}\begin{eqnarrayabc}#2\end{eqnarrayabc}}\newline}
\ncm{\eqbox}[2]{\newline\parbox{1.5cm}{#1}\hfill
  \parbox{12cm}{\beanon #2\eeanon}\hfill
  \parbox{1cm}{\bea\eea}\newline}
\ncm{\nr}[1]{\parbox{1cm}{\begin{eqnarrayabc}#1\end{eqnarrayabc}}\\}

\ncm{\kal}[1]{\mbox{$\cal #1 $}}
\ncm{\mrk}[1]{\!\!\! #1 \!\!\!} 
\ncm{\qed}{\hspace*{0.4cm}\rule{0.24cm}{0.24cm}}  
\ncm{\mbold}[1]{\mbox{\boldmath $ #1 $}}   
\ncm{\bm}{\mbold}
\ncm{\str}{\stackrel}
\ncm{\sub}{\subset}
\ncm{\e}{\varepsilon}
\ncm{\ka}{\kappa}
\ncm{\inputc}[1]{\begin{center}\input{#1}\end{center}}
\ncm{\lto}{\longrightarrow}
\ncm{\x}{\times}
\ncm{\bmm}{\bm{\cal M}}
\ncm{\cp}{{\bf P}}    
\ncm{\bfp}{{\bf P}}
\ncm{\bmi}{\bm{i}}
\ncm{\bmom}{\bm{\om}}
\ncm{\bmOm}{\bm{\Om}}
\ncm{\res}{\restriction}
\ncm{\bmL}{\bm{\cal L}}
\ncm{\bmell}{\bm{\ell}}
\ncm{\bmE}{\bm{\cal E}}
\ncm{\bme}{\bm{e}}
\ncm{\bmpi}{\bm{\pi}}
\ncm{\bmr}{\bm{r}}
\ncm{\bmsigma}{\bm{\sigma}}
\ncm{\wt}{\widetilde}
\newcommand{\beaa}{\begin{eqnarray}}
\newcommand{\eeaa}{\end{eqnarray}}

\begin{document}

\author{{\bf Oscar Bolina}\thanks{Supported by FAPESP under grant
01/08485-6. {\bf E-mail:} bolina@if.usp.br} \\
Departamento de F\'{\i}sica-Matem\'atica\\
Universidade de S\~ao Paulo\\
Caixa Postal 66318 S\~ao Paulo\\
05315-970 Brasil\\
}
\title{\vspace{-1in}
{\bf Trotter formula and thermodynamic limits}}
\date{}
\maketitle
\begin{abstract}
\noindent
We discuss the interchangeability of the thermodynamic limit 
$\beta \rightarrow \infty$ and the infinite limit of the Trotter 
number $n \rightarrow \infty$ when Trotter formula $e^{H_{0}+V}
=\lim_{n \rightarrow \infty} (e^{H_{0}/n} e^{V/n})$ is used to 
calculate partition functions with Hamiltonians of the form 
$H=H_{0}+V$. 

\noindent
{\bf Key words:}  Trotter Formula, Thermodynamic Limit \hfill \break
{\bf PACS numbers:} 05.30.-d.
\end{abstract}
\noindent
This note deals with thermodynamic limits with respect of parameters in 
certain thermodynamic functions of model Hamiltonians given by a sum of
non-commuting operators $H=H_{0}+V$. The first parameter is the inverse
temperature $\beta$ that appears, for instance, in the definition of the
free energy 
\begin{equation} \label{FE1}
E=\lim_{\beta \rightarrow \infty} -\frac{1}{\beta} \ln Z
\end{equation}
where $Z$ is the partition function of a particular model considered.
\newline
It is sometimes advantageous to write the partition function of 
quantum models in the Trotter representation by the use of the
Trotter formula (for pair of operators limited from below)
\beq\label{T}
e^{-\beta (H_{0}+V)}=\lim_{n \rightarrow \infty} ( e^{-\beta H_{0}/n}
e^{-\beta V/n})^{n}.
\eeq
This representation is specially useful in quantum mechanics 
since the operators $H_{0}$ and $V$ will not in general commute, 
and one will not be allowed to decompose the exponential operator
into a product of exponentials. The Trotter formula achieves 
this at the expense of formally transforming the quantum problem 
into a classical one having one extra dimension. As a result, a
second parameter {\it n} introduced, so that the ratio ${\beta}/n$ 
plays the role of a lattice spacing in the extra "temperature" 
direction $[0, \beta]$. This involves taking another limit in 
the free energy function
\begin{equation}\label{FE2}
E=\lim_{\beta \rightarrow \infty} \lim_{n \rightarrow \infty} 
-\frac{1}{\beta} \ln Tr ~ T^{n}
\end{equation}
where $Tr ~ T^{n}$ is the trace of the $n^{th}$ power of transfer 
matrix $T$.
\newline
The order of the limits in (\ref{FE2}) can not in general be interchanged.
It is not always clear whether it is possible to interchange 
the limits even in simple problems and an analysis of its validity
is beyond our scope here.
\newline
We want to point out that $n$ and $\beta$ always appear as a single
variable in the ratio $\beta/n$ throughout the calculation and the
question of the order of the limits is relevant only at the end, when
the effect of letting $n \rightarrow \infty$ first definitely
manifests itself. 
\newline
In order to illustrate the role played by the parameters $\beta$ and 
{\it n} in the thermodynamic functions and see how the {\it  
n-to-infinity-first} limit imposes itself when interchangeability fails 
it suffices to consider the free energy of a simple spin model Hamiltonian
$H=\sigma_{z}+\lambda \sigma_{x}$, where $\lambda$ is a constant and
$\sigma_{z}$ and $\sigma_{x}$ are the usual Pauli spin matrices
\[
\sigma_{z}=\left [
\begin{array}{cc}
1 & 0 \\
0 & -1 \\
\end{array}
\right ], \;\;\;\;\;
\sigma_{x}=\left [
\begin{array}{cc}
0 & 1 \\
1& 0 \\
\end{array}
\right ]
\]
For this model, a direct evaluation of the partition function $Z=Tr
e^{-\beta H}$ in (\ref{FE1}) is available since the eigenvalues of the
Hamiltonian are $E_{\pm}=\pm \sqrt{1+\lambda^{2}}$. Thus the trace can
be written as a sum over  the energy eigenstates $Z=e^{-\beta
E_{-}}+ e^{-\beta E_{+}}$. The analysis is facilitated when we write  
$Z=e^{-\beta E_{-}}(1+e^{-\beta(E_{+}-E_{-})})$, since in the limit
$\beta \rightarrow \infty$ the lower eigenvalue term dominates, and 
the relative error goes to zero exponentially. In this limit the free
energy (\ref{FE1}) becomes
\beq\label{AV}
E=\lim _{\beta \rightarrow \infty} -\frac{1}{\beta} 
\ln e^{-\beta E_{-}}(1+e^{-\beta(E_{+}-E_{-})})
=-\sqrt{1+\lambda^{2}}.
\eeq
The Trotter formula approach to this problem consists in applying
(\ref{T}) to the partition function of the model. This leads to the
transfer matrix formalism in which the partition function reads
$Z=\lim_{n \rightarrow \infty} \sum_{\sigma_{z}} \langle \sigma_{z}
~T^{n} ~\sigma_{z} \rangle = \lim_{n \rightarrow \infty} Tr ~ T^{n}$,
where {\it T} is the $2 \times 2$ transfer matrix
\[
T=e^{-\frac{\beta}{n}}  \cosh{\frac{\beta \lambda}{n}}
\left [
\begin{array}{cc}
1 & -\tanh{\frac{\beta \lambda}{n}}\\
-e^{2\frac{\beta}{n}} ~\tanh{\frac{\beta \lambda}{n}} & 
e^{2\frac{\beta}{n}} \\
\end{array}
\right ],
\]
with eigenvalues 
\[
\lambda_{\pm} = \cosh{\frac{\beta}{n}}  \cosh{\frac{\beta \lambda}{n}} 
\pm \sqrt{\cosh^{2}\frac{\beta}{n} \cosh^{2}\frac{\beta \lambda}{n}-1 }.
\]
Although so far $\beta$ and {\it n} have always appeared together as 
the ratio ${\beta}/{n}$, from this point on we will see that in order
to obtain the correct asymptotic value (\ref{AV}) we have to let $n
\rightarrow \infty$, with a finite $\beta$.
\newline
In the first place the eigenvalues of $T^{n}$ are the $n^{th}$
power of the eigenvalues of $T$, so that $Tr ~ T^{n}=\lambda^{n}_{+}
(1+{\lambda^{n}_{-}}/{\lambda^{n}_{+}})$ and we could be tempted to
proceed as in the ordinary transfer-matrix method by taking only the
maximum eigenvalue. However, as a note of caution on the effect of
{\it n}, we must point out that the eigenvalues depend on {\it n}
and the ratio $({\lambda_{-}}/{\lambda_{+}})^{n}$ may remain finite
and therefore may not be negligible even when {\it n} is infinite.
\newline
In our case, $({\lambda_{-}}/{\lambda_{+}})^{n}$ {\it does} remain 
finite when $n \rightarrow \infty$ and converges to $\exp\{-2 \beta
\sqrt{1+\lambda^{2}}\}$. Only after this check the free energy becomes 
\beq\label{EI}
E= \lim _{\beta \rightarrow \infty} \lim _{n \rightarrow \infty}
-\frac{1}{\beta} (n \ln{\lambda_{+}} +
\ln{(1+e^{-2\beta\sqrt{1+\lambda^{2}}}}))  
\approx \lim _{\beta \rightarrow \infty} \lim _{n  \rightarrow \infty}
-\frac{n}{\beta} \ln \lambda_{+}
\eeq
and we can disregard the second factor in comparison with
$n\ln{\lambda_{+}}$.
\newline
Another manifestation of the {\it n-to-infinite-first} limit is that
it is the first order term resulting from the expansion of (\ref{EI}) 
in power series of $\beta/n$, {\it when $\beta/n \ll 1$}, that yields 
the correct free energy (\ref{AV}) 
\[
E= -\frac{n}{\beta} \ln(1+\frac{\beta}{n} \sqrt{1+\lambda^{2}}) 
= -\sqrt{1+\lambda^{2}}. 
\]
The above considerations stem from the author's tentative to solve an open
problem in statistical mechanics. This is the problem of proving the 
existence of a Kosterlitz--Thouless phase in the ground state ($\beta 
\rightarrow \infty$) of a one-dimensional array of quantum rotators
(See \cite{BP} for a definition of the model and some results towards
this proof). One apparent very natural approach to this problem consists
in using the Trotter representation to map the one-dimensional quantum
rotator system into a two-dimensional classical system in the form we have
stated above. Since a Kosterlitz--Thouless phase has already been proven
for the two-dimensional classical rotator \cite{FS}, it is a generally
held opinion that the quantum rotators also exhibit this transition, but 
the proof is still missing.

\end{document}